\documentclass[a4paper,11pt]{article}
\pdfoutput=1 

\usepackage{jinstpub} 

\usepackage{array}
\usepackage{graphicx}      
\usepackage{tabularx}          
\usepackage[utf8]{inputenc}
\usepackage{array}
\usepackage{makecell}

\title{\boldmath Material Discrimination in Cosmic Muon Imaging using Pattern Recognition Method}

\author[a,b,1]{Sridhar Tripathy\note{Corresponding author}}
\author[a,b]{Jaydeep Datta}
\author[a,b]{Nayana Majumdar}
\author[a,b]{and Supratik Mukhopadhyay}

\affiliation[a]{Saha Institute of Nuclear Physics,\\AF Block, Sector 1, Salt Lake, Kolkata 700064, India}
\affiliation[b]{Homi Bhabha National Institute, \\Training School Complex, Anushaktinagar, Mumbai 400094, India}

\emailAdd{sridhar.tripathy@saha.ac.in}

\abstract{ This work reports numerical simulations and  statistical tests carried out to study the performance of a prototype imaging setup to be developed for material discrimination using Muon Scattering Tomography (MST). The reconstructed images have been processed with a Pattern Recognition Method (PRM) to discriminate the low-Z and high-Z materials. The same method has been applied to examine the performance of the setup in recognizing the shape of the test objects. Design parameters of the setup, such as, area of the muon detectors, their physical placement,  along with detector performance (spatial resolution, etc.) have been discussed. To evaluate the efficacy of the newly proposed PRM, the results have been compared to another technique based on cluster finding.}


\arxivnumber{2003.05504 } 

\begin{document}
\maketitle
\flushbottom

\section{Introduction}
\label{sec:intro}

In order to accomplish non-destructive imaging of any object, tomography utilizing the scattering of cosmic muons is one of the widely used techniques. It is found useful particularly in those cases where discrimination between materials with high and low atomic numbers~(Z) is required. A couple of examples may be imaging of spent nuclear fuel container~\cite{Fuel} and detection of contraband fissile materials~\cite{Borozdin}. The technique relies upon repeated elastic Coulomb scattering suffered by the muons while passing through an object due to the electric field of the atomic nuclei therein. These collisions individually result in angular deflections which can be obtained following the Rutherford's formula. For a thicker~/~denser object there can be multiple scatterings~($>20$). For small angles, it can be described by a Gaussian distribution~\cite{Bethe},~\cite{Highland},~\cite{lynch}. The standard deviation~($\sigma$) of the distribution is given by,
	\begin{equation}
	\label{eq:scatt}
	\sigma = {13.6 \over \beta p}{\sqrt {L \over X_0}}\Big(1 + 0.038 ln{L \over X_0}\Big)
	\end{equation}
where $p$ is the momentum of the muon, $L$ is the distance traversed by it in the object and $\beta$ is the ratio of its velocity~($v$) to the velocity of light~($c$). The radiation length, $X_0$~(cm), is a property of the object material which is related to its atomic weight,~$A$, atomic number,~$Z$, and density, $\rho$~(g$.$cm$^{-3}$), in the following way.
	\begin{equation}
	\label{eq:rL}
	X_0 = {716.4 A \over {\rho Z(Z+1) ln\Big({287 \over {\sqrt Z}}\Big)}}
	\end{equation}
From equations \ref{eq:scatt} and \ref{eq:rL}, it is observed that the width of the scattering angle distribution has a strong dependence on the density and atomic number of the object material. Therefore, an estimate of the scattering angle can lead to identification of materials for which tracing of muon trajectories before and after its interaction with the object with reasonable accuracy is required.
 
The authors plan to construct a small-scale prototype imaging setup for studying the application of MST technique in discriminating materials with different Z number, especially in imaging of civil structures. A model of the setup, containing six position sensitive gaseous ionization detectors for tracing the muon trajectories above and below a region of interest~(ROI) has been constructed in Geant4~\cite{Gea}. The setup has been subjected to a limited exposure of the cosmic muon flux received at the sea level using Cosmic Ray Library~(CRY)~\cite{Cry}. The hits obtained from the detectors have been used for reconstruction of the incoming and outgoing tracks for each event. The fitting of the tracks and determining the scattering vertices have been done using Binned Clustering Algorithm~(BCA)~\cite{Thom} followed by calculation of the angle. The description of the process is also available in another report of the authors where a separate method by material discrimination has been adopted~\cite{Jinst_ST}.
The tracks with scattering angle below a certain threshold value have been rejected, which are mostly due to particles which cross the ROI, yet don't pass through any target object (un-scattered muons) and high-energy muon which scatter feebly. Tracking is better for detectors with precise spatial resolutions, however obtaining detectors with fine resolution requires a dear expense for electronics and data acquisition system. In section~\ref{sec:ImPer}, a $t$-statistics test has been conducted on images obtained for different spatial resolutions of the detector with the one using ideal resolution as reference.

Analyses based on scattering vertices and the scattering angle and related variables have been shown to detect the presence of high-Z materials with rational precision in a few published reports~\cite{Thom},~\cite{schultz}. However, identification of multiple targets with dissimilar properties in a given ROI requires analysis of the whole three-dimensional volume. In the present work, it has been proposed that the orthogonal planes i.e.~XY, XZ and YZ planes can be analyzed to investigate material content, location and dimension of the target object(s) instead of scanning the ROI volume. Two-dimensional images are obtained by the projection of scattering vertices in the orthogonal planes. Then the material discrimination capability of the two-dimensional imaging technique has been tested. Images obtained using simulation of samples of known materials, are used as a database. A test image with several materials has been investigated, a $t$-test compares the parameters of the test image with ones in the given database.    

The statistical tests can be used to differentiate materials, however the dimension and shape of the object can't be accurately predicted. Therefore, a pattern recognition algorithm has been proposed to discriminate high-Z materials from the low-Z materials and to identify their physical profile in a unknown ROI based on the learning parameters obtained beforehand from the images of known samples. It's description has been provided in section~\ref{sec:ImProc}. The results obtained using this method have been compared with a clustering algorithm, namely, Density-Based Spatial Clustering of Applications with Noise~(DBSCAN)~\cite{Dbsc},~\cite{Riggi}. The software analysis in this work has been done using ROOT~\cite{ROOT} and MATLAB~\cite{Mat}. 

\section{Event Generation \& Track Reconstruction}
\label{sec:evengen}

The simulation processes involved in generating the cosmic muon events, geometry of the prototype imaging setup and the subsequent reconstruction of the tracks from the muon hits registered in the detectors have been discussed here.  

\subsection{Particle Generation}
The CRY has been used as the particle generator in the Geant4 simulation to produce muon flux for the latitude and altitude of the experimental laboratory. A normal day other than the solar maxima or minima has been considered in the generator along with a fixed exposure time of 3.5 hours. FTFP\_BERT physics list~\cite{FTFP} has been used to govern the tracking of particles. It takes care of all the electromagnetic and weak interactions of the charged particles, and gamma radiations. The hit information from each of the detectors have then been extracted for track reconstruction.
\subsection{Setup Geometry}
\begin{figure}[h]
	\centering
	\includegraphics[trim = 0 0 10 10, clip, angle = 0, width=0.5\textwidth]{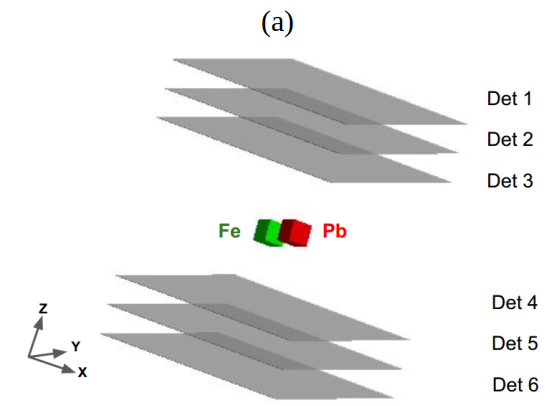}
	\enspace
	\includegraphics[width=0.45\textwidth,height=0.40\textwidth]{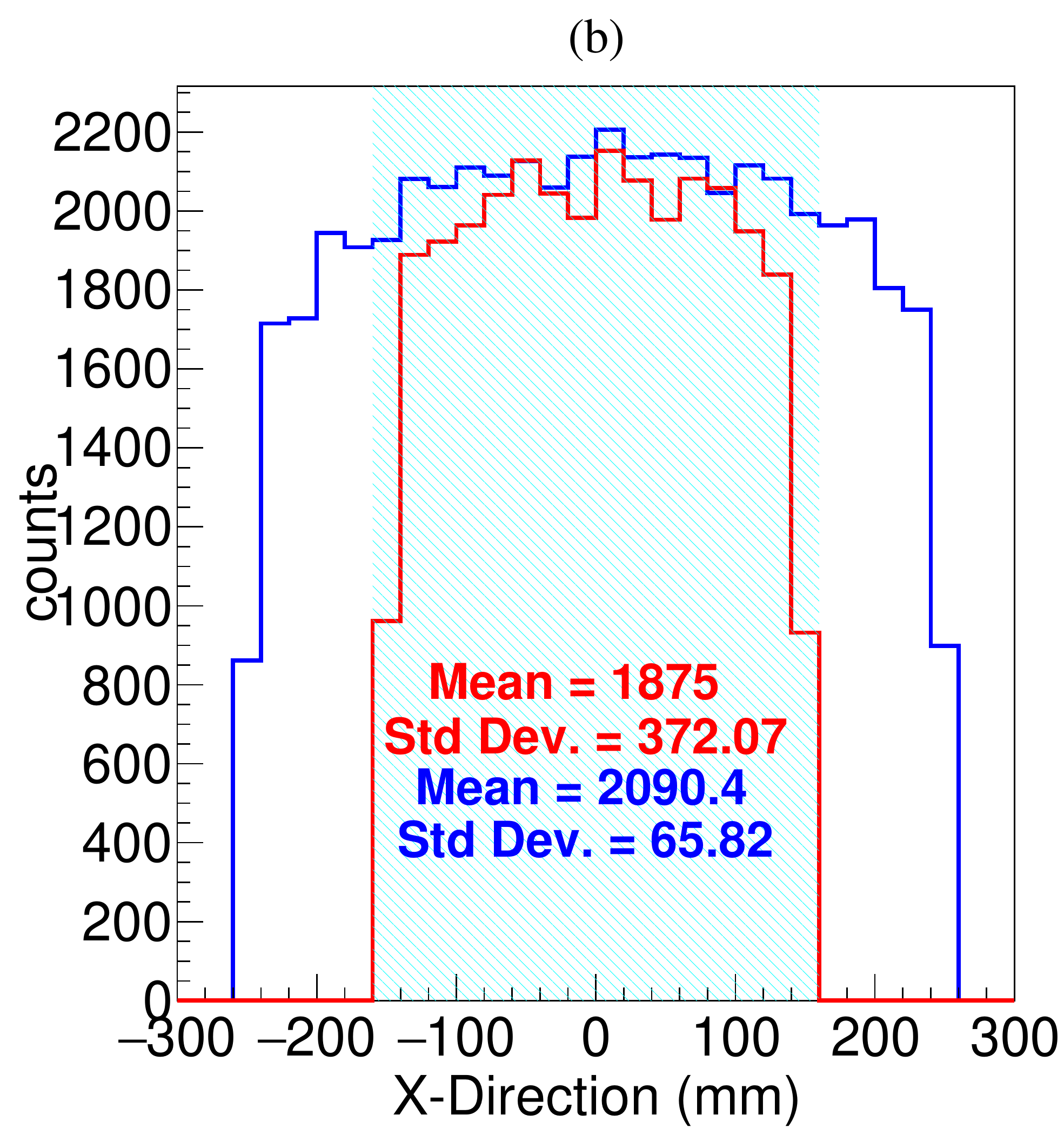}
	\caption{\label{setup}(a) Schematic representation of the setup, (b) Muon hits received on the lowermost detector along the X direction for detectors with area 30~cm~$\times$~30~cm~(red) and  50~cm~$\times$~50~cm~(blue). The detectors were placed with a separation of $d$ = 7cm. The shaded portion in cyan represents the ROI.}
\end{figure}

The schematic presentation of the imaging setup, as modeled in Geant4, is shown in figure~\ref{setup}~(a). Two sets of tracking detectors, each comprising of three gaseous detectors, have been placed on either sides of an ROI of dimension 30~$\times$~30~$\times$~30~cm$^3$. Each of the detectors has been designed with a gaseous volume of 2~mm thickness bound by two parallel electrodes. As the focus of this work is obtaining the hit from the detector volumes, only the coulomb scattering and ionization interaction points have been taken as valid muon hit inside the detector volume. The signal formation from the avalanche of secondary electrons have not been considered in the simulation model,  Two design parameters, the area of the detectors and the vertical separation between them have been varied to maximize the efficiency of the setup within the limit of practical feasibility. To examine the uniformity of the muon hits in the vertical direction, the hit patterns received for two different detector sizes have been analyzed. The figure~\ref{setup}(b) displays the mean values and standard deviations of the hit patterns received on the bottom-most detector along the X direction covering the ROI. For the larger area~(50~$\times$~50 cm$^2$), mean value of the hit distribution is higher while the standard deviation is less, which result in better uniformity with higher yield. Therefore the dimension of each detector has been chosen to be 50~$\times$ 50~cm$^2$. For optimizing the arrangement of the detectors, some parameters relevant in governing muon detection efficiency of the system, such as, i) acceptance angle~($\phi$), ii) angular resolution~($\delta \theta$), and iii) number of muon events registered in each detector in one hour~($N$),  have been calculated for different values of their separation~($d$) in Z direction. The results obtained for the detector area 50~$\times$~50~cm$^2$ have been tabulated in table~\ref{set}. It can be followed that~$\delta \theta$ improves if $d$ increases. However, it leads to a deterioration in~$\phi$, and the number of events registered, $N$. The effect of~$N$ and $\phi$ on the present goal is more vital than that of $\delta \theta$. Moreover, considering the practical issues of mechanical handling the detector separation was chosen 7~cm. For this separation, the intensity received by the detector setup is about 74$\%$ of the total muons generated.
\begin{table}
	\centering	
\begin{tabular}{|p{2cm}|p{2cm}|p{2cm}|p{2cm}|}

\hline
d (cm) & $\phi$ (deg) & $\delta$$\theta$ (deg) & N (in 1 hr)\\
\hline
7 & 40.76 & 0.86 & 38949\\
\hline
10 & 35.54 & 0.72& 34602\\
\hline
15 & 29.05 & 0.56 &21931\\
\hline
20 & 24.44 & 0.46 & 16156\\\hline

\end{tabular}
	\caption{\label{set} Muon detection parameters of the imaging setup.}
\end{table}
\subsection{Track Reconstruction}
The hits produced in the sensitive region of all six detectors have been stored in a root file. In real life experiment, the hits due to muons or other particles/radiation can not be distinguished in tracking detectors. Therefore to emulate experimental scenario, besides $\mu^+$ and $\mu^-$ hits, $e^-$, $e^+$ and $\gamma$ interactions have also been stored. The extra noise thus induced has been removed using track-selection criteria. 
Following are the selection criteria implemented: 
i) A track must have hits on all the detectors, ii) the angle between the tracks reconstructed from the pair of hits on the first and the second detectors and that from the pair of the first and the third ones must be less than 0.5$^{\circ}$, iii) the angle between the tracks in upper and lower sets of detectors must not be less than 1.5$^{\circ}$. Following the reconstruction of the incoming and outgoing trajectories for each muon event, the probable scattering vertex has been estimated using the BCA method. The uncertainty in the hit position has been introduced by generating a random value with the actual hit point as its mean and the standard deviation equal to the spatial resolution intended. For the sake of simplicity, the spatial resolution of all the detectors has been considered same. Following the determination of the vertex, the scattering angle between the two concerned trajectories has been determined. 

\section{Image Formation \& Statistical Test}
\label{sec:ImPer}
Two-dimensional images have been obtained projecting the scattering vertices inside the ROI, on orthogonal planes i.e. XY, XZ and YZ planes. Here the XY projection has been used for analysis. The performance of the setup for producing images has been evaluated with statistical significance for variation in detector resolution and discrimination between materials using $t$-statistics. The $t$-value has been calculated in each case using equation~\ref{t1}. 

\begin{equation}
\begin{split}
\label{t1}
t=\frac{\mu_{1}-\mu_{2}}{s_{v}\left[\frac{1}{n_{1}}+\frac{1}{n_{2}}\right]}\\
\textrm{with}\;\;\;
s_{v}=\sqrt{\left[\frac{(n_{1}-1){s_{1}}^2+(n_{2}-1){s_{2}}^2}{n_{1}+n_{2}-2}\right]}
\end{split}
\end{equation}
where $\mu_{1}$, $n_{1}$ and $s_{1}$ are the sample mean, sample size and standard deviation of the reference distribution while $\mu_{2}$, $n_{2}$ and $s_{2}$ are the corresponding parameters of the test data set. The test begins with assuming the null hypothesis that the data from the two images are same, and the significance level~($\alpha$) has been set to be 0.05. The $p$-value for each case has been estimated from the $t$-statistics. 
\begin{figure}[h]
	\centering 
	\includegraphics[width=.9\textwidth,height=.9\textwidth]{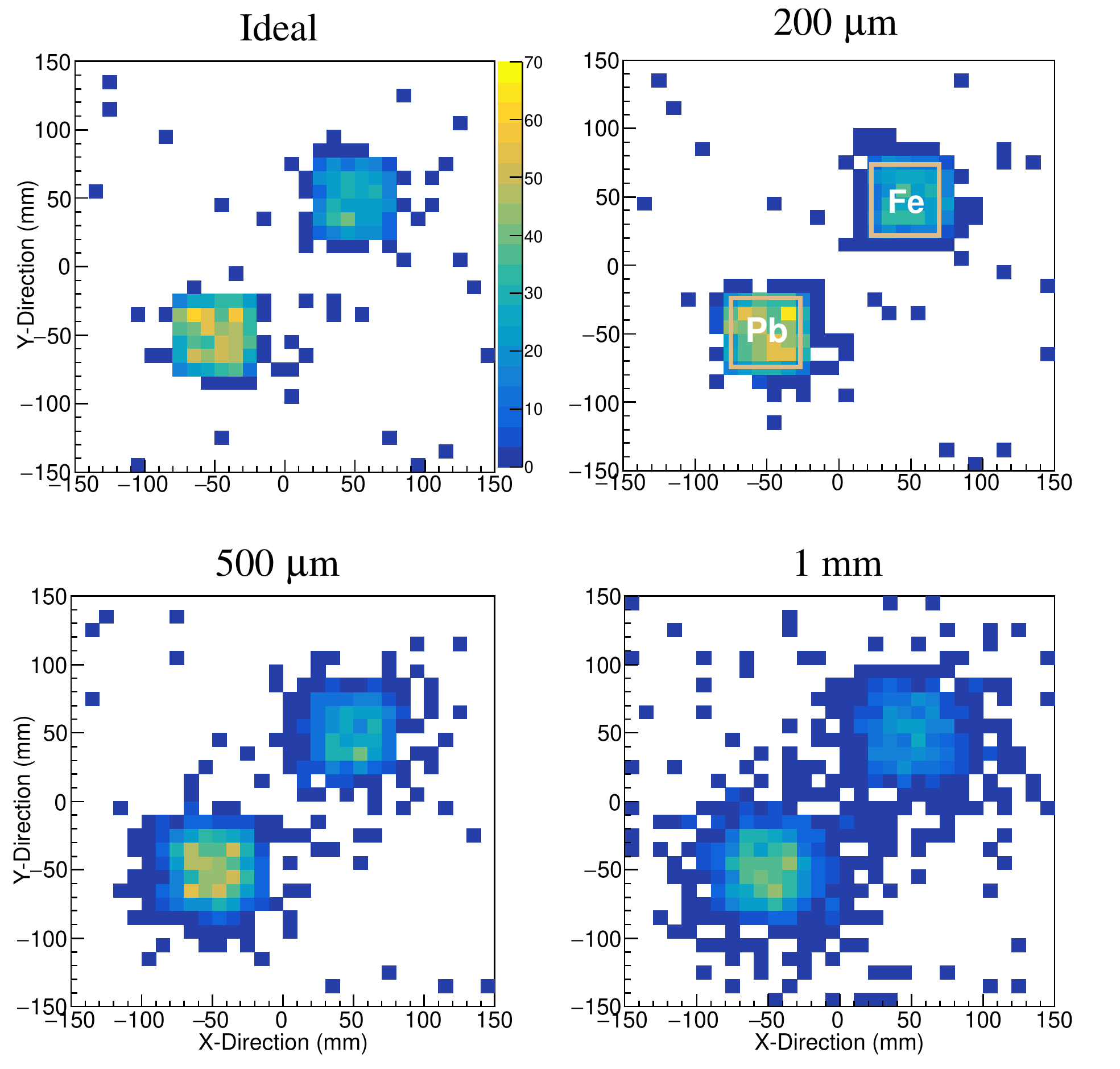}
	\caption{\label{scver_xy} XY images for the ideal resolution, 200 $\mu$m, 500 $\mu$m and 1 mm. The object boundaries have been marked by solid line in the second plot.}
\end{figure}

\begin{table}[h]
	\centering
	\smallskip	\begin{tabular}{|c|c|c|}
		\hline
		Test&$t$-value&$p$-value\\
		\hline
		Ideal vs 200 $\mu$m &-0.37&\textbf{0.71}\\
		\hline
		Ideal vs 500 $\mu$m &-1.43& \textbf{0.15}\\
		\hline
		Ideal vs 1 mm &-3.60&8e-4 \\
		\hline
	\end{tabular}
	\caption{\label{ttest_xy} Comparison of $t$-values and $p$-values obtained from the tests between XY images obtained with different spatial resolutions and the ideal resolution. The $p$-values in bold indicate that the null hypothesis can't be rejected.}
\end{table}

\subsection{Variation in Detector Resolution}

In the test model, two cubes made up of materials with moderately low-Z~(iron) and high-Z~(lead) values, each of dimension 5 cm, have been kept such that their center lies on the central horizontal~(XY) plane of the ROI. The XY images obtained are shown in figure~\ref{scver_xy} for three different resolutions of the detector: 200~$\mu$m, 500~$\mu$m and 1~mm along with the ideal one. The results of the statistical test have been tabulated in table~\ref{ttest_xy} where the image obtained with ideal resolution has been regarded as the reference. It can be noted that the data with spatial resolution of 200 and 500 $\mu$m are not significantly differentiable from the ideal case because the null hypothesis can not be rejected for these resolutions as indicated by their $p$-values. Therefore, these two resolutions have been considered for all the later studies.

\begin{figure}[h]
	\centering 
	\includegraphics[width=0.90\textwidth,height=0.42\textwidth]{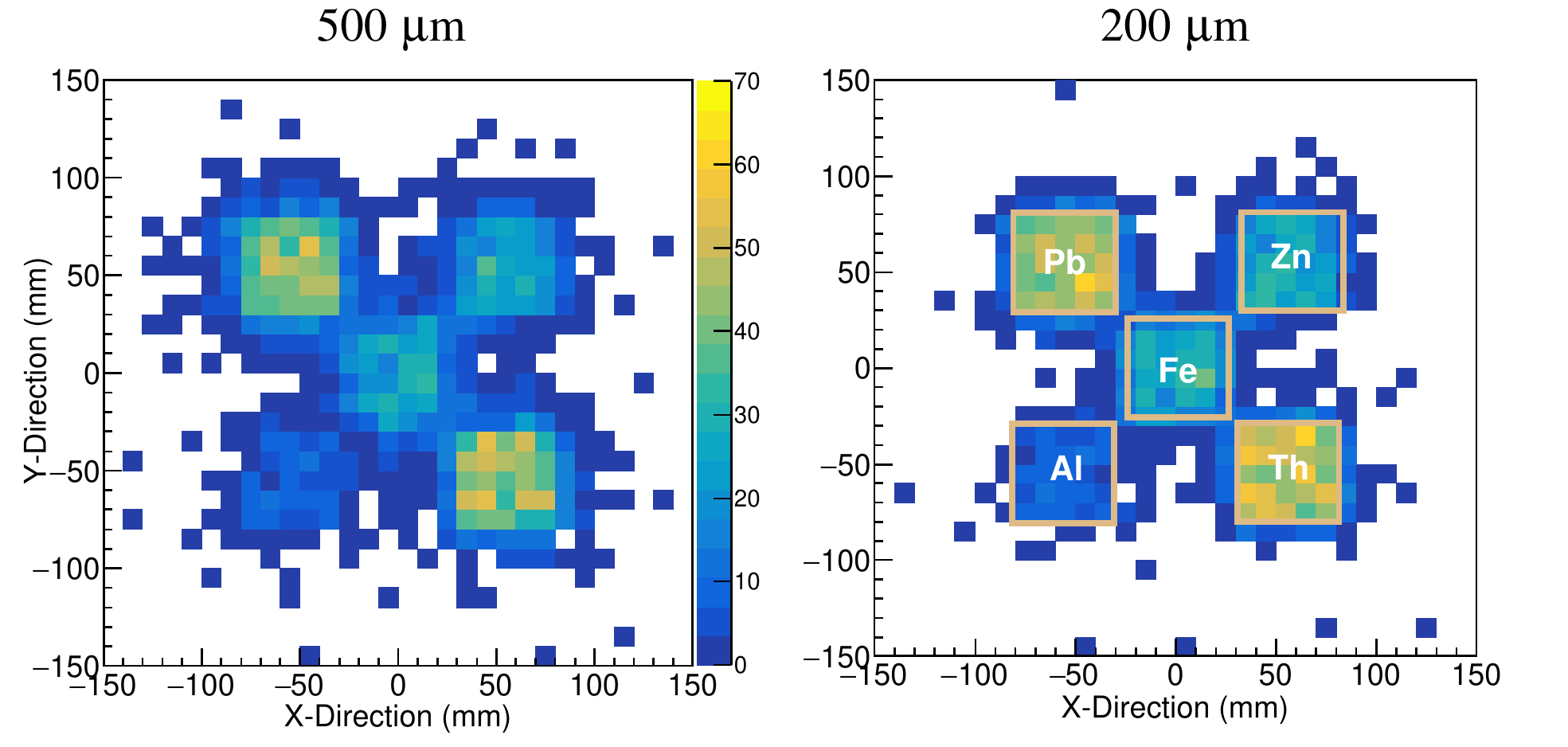}
	\caption{\label{sch} XY images of scattering vertices for spatial resolutions of 200 $\mu$m and 500 $\mu$m. The object boundaries have been marked by solid line in the second figure.}
\end{figure}
\begin{table}[h]
	\centering
	\begin{tabular}{|c|c|c|c|c|c|}\hline
		\multicolumn{6}{|c|}{\textbf{500 $\mu$m}}\\\hline
		\multicolumn{3}{|c}{W.r.t Fe} &
		\multicolumn{3}{|c|}{W.r.t Pb}\\\hline
		test  & $t$-value & $p$-value &test  & $t$-value & $p$-value \\ \hline
		Fe vs Pb & 12.575 &   4.71e-12        & Pb vs Pb                 & 0.223                 & \textbf{0.825}        \\ 
		Fe vs Zn & 1.219 & \textbf{0.235}         & Pb vs Zn                 & -10.469                 & 1.99e-10        \\ 
		Fe vs Fe & 0.212 & \textbf{0.834}         & Pb vs Fe                 & -6.697                 & 6.31e-7        \\ 
		Fe vs Th & 11.074 & 6.47e-11         & Pb vs Th                 & 0.254                & \textbf {0.802}        \\ 
		Fe vs Al & -12.817 & 3.15e-12         & Pb vs Al                 & -22.591                 & 1.10e-17        \\  \hline
		\multicolumn{6}{|c|}{\textbf{200 $\mu$m}}\\\hline
		\multicolumn{3}{|c}{W.r.t Fe} &
		\multicolumn{3}{|c|}{W.r.t Pb}\\\hline
		test  & $t$-value & $p$-value &test  & $t$-value & $p$-value \\ \hline
		Fe vs Pb & 14.705 &   1.67e-13        & Pb vs Pb                 & -0.467                 & \textbf{0.645}        \\ 
		Fe vs Zn & 1.030 & \textbf{0.313}         & Pb vs Zn                 & -12.584                 & 4.64e-12        \\ 
		Fe vs Fe & -0.106 & \textbf{0.917}         & Pb vs Fe                 & -6.862                 & 4.26e-7        \\ 
		Fe vs Th & 15.515 & 5.19e-14         & Pb vs Th                 & 0.499                & \textbf {0.622}        \\ 
		Fe vs Al & -15.538 & 5.03e-14         & Pb vs Al                 & -24.15                 & 2.39e-18        \\  \hline
	\end{tabular}
	\caption{\label{tt} Comparison of $t$-values and $p$-values between images of different materials and that of iron and lead as references with detector resolutions 200~$\mu$m and 500~$\mu$m. The $p$-values in bold are higher than $\alpha$ and hence the null hypotheses can not be rejected for those cases.}
\end{table}
\subsection{Material Discrimination}
Five cubes of materials with Z-values ranging from low to high, such as, aluminium (Z = 13), iron (Z = 26), zinc (Z = 30), lead (Z = 82) and thorium (Z = 90) of dimension 5 cm have been placed with their center lying on the central plane of the ROI. The $t$-test has been done to compare the discrimination of different materials from either of the iron and lead which represent moderately low and high Z-values respectively as mentioned earlier. The test has been performed for both the detector resolution of 200 and 500 $\mu$m. The results have been tabulated in table~\ref{tt} which show that the test has rejected materials with different Z-number and has not found evidence to differentiate materials close in Z-number (as close as 4 to 8) as indicated by the $p$-values. These results encourage the use of an two dimensional image processing algorithm to discriminate materials on the basis of muon imaging. 
\section{Image Processing}
\label{sec:ImProc}
Besides identification and discrimination of the materials, it is imperative to extract information about the physical structure (shape~/~dimension) of the objects. These can be predicted by analyzing the orthogonal images obtained from track reconstruction, although the competence of image processing is considerably restricted by the spatial resolution of the detectors. A step-wise algorithm based on pattern recognition technique has been followed to process the projected images. The description of the PRM has been given below.
 
\begin{figure*}[htbp]	
	\centering 
	\includegraphics[trim = 80 20 80 20, clip, angle = 0,width=0.9\textwidth]{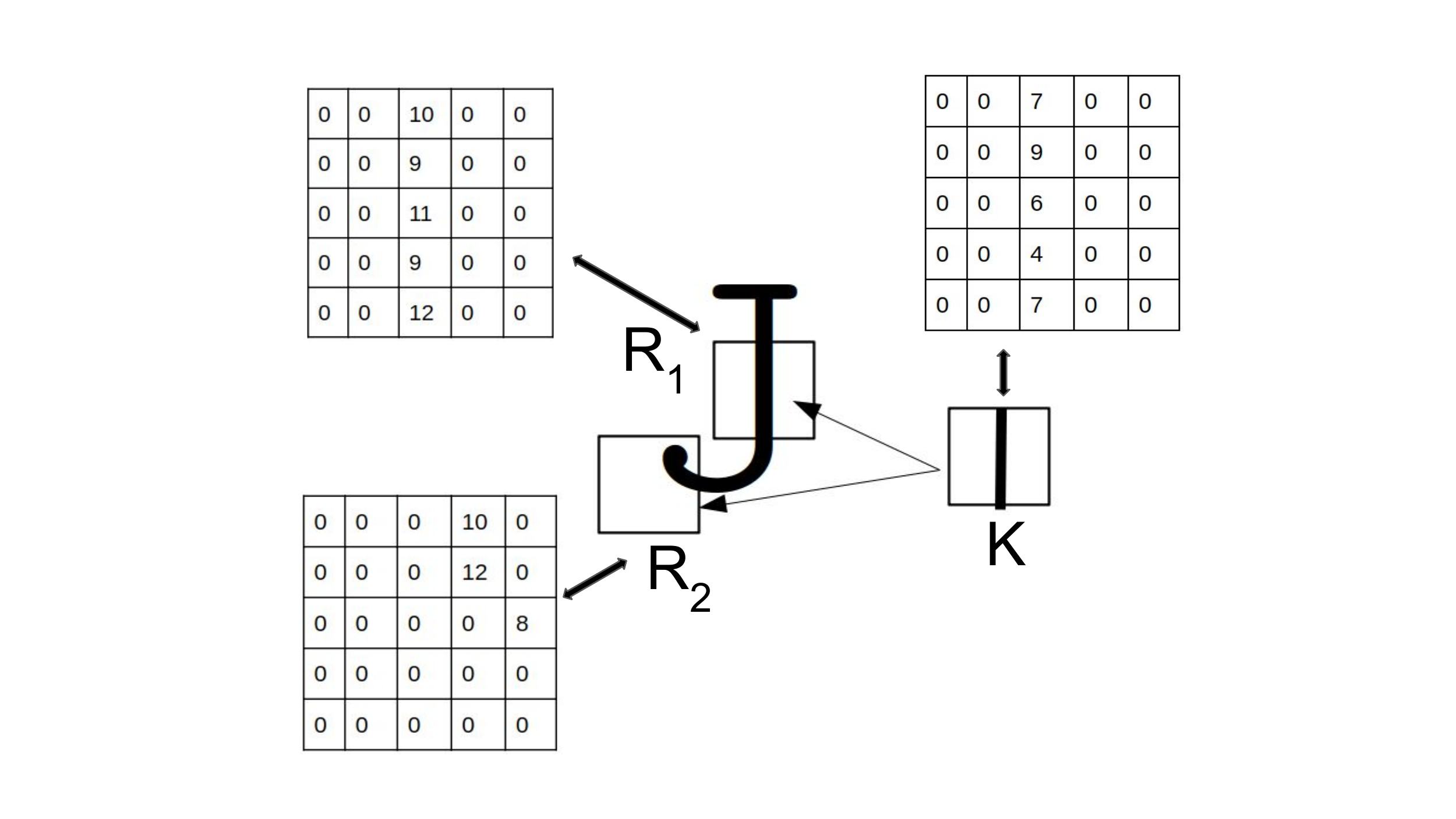}
	\caption{\label{cnn} A filter `K' searches for a similar pattern `R$_{1}$'.}
\end{figure*}

A filter is designed based on a specific pattern of a sample image, and this filter is used to scan the test objects for the same pattern. To explain the method, an example has been depicted in figure \ref{cnn} where a filter `K' (called kernel) searches for a physical pattern like a straight line in two receptive fields, `R$_1$' and `R$_2$' in the letter `J'. The representative matrices of `K' and the selected regions `R$_1$', `R$_2$', have been displayed. The elements of these matrices are the numerical entries to represent image pixels. The results of the multiplication of `R$_1$' and `R$_2$' to `K' are given below in \ref{RK}. It indicates that a similar pattern of the filter `K' is present in the receptive field, `R$_1$' as its multiplication to `K' has produced a significantly large value. Therefore, the algorithm as designed here, has performed like a high-pass filter that allows the range of values lying above the attributes it is designed with.

\begin{equation}
\label{RK}
\begin{split}
\textrm{R$_1$*K :}~7*10+9*9+6*11+4*9+7*12 = 337
\\
\textrm{R$_2$*K :}~7*0+9*0+6*0+4*0+7*0 = 0
\end{split}
\end{equation}

The images produced by the setup have been analyzed using the PRM, as described above, to discriminate or approve materials with representative value of a pattern similar or larger than that set by the filter. A parameter, named as clustering density, $\rho_c$, has been defined to represent the number of scattering vertices inside a given pixel of the image. Since $\rho_c$ has been found to increase with the Z-values of the material, it has been opted as a test parameter for material discrimination. The images have been segmented into 75 $\times$ 75 pixels, each having an area of 4 $\times$ 4 mm$^2$. In the present work, two filters of size 4~$\times$~4 have been designed with clustering densities obtained for Z-values of iron and lead. These two materials have been particularly chosen to represent low and high Z-values. It can be mentioned here that any other material can be used to design a filter. The size of the pixels and size of the filter matrix are decided based on computation time, required accuracy in shape identification, amount of data etc. The relevant filter is multiplied to the submatrices of same dimension of the image matrix. If the result of the matrix multiplication is found higher than or equal to the threshold for a given material, the corresponding submatrices are marked positive. Next the algorithm shifts by one stride and continues the process until the whole matrix scan is complete.
\begin{figure}[htbp]
	\centering 	
	\includegraphics[width=.9\textwidth]{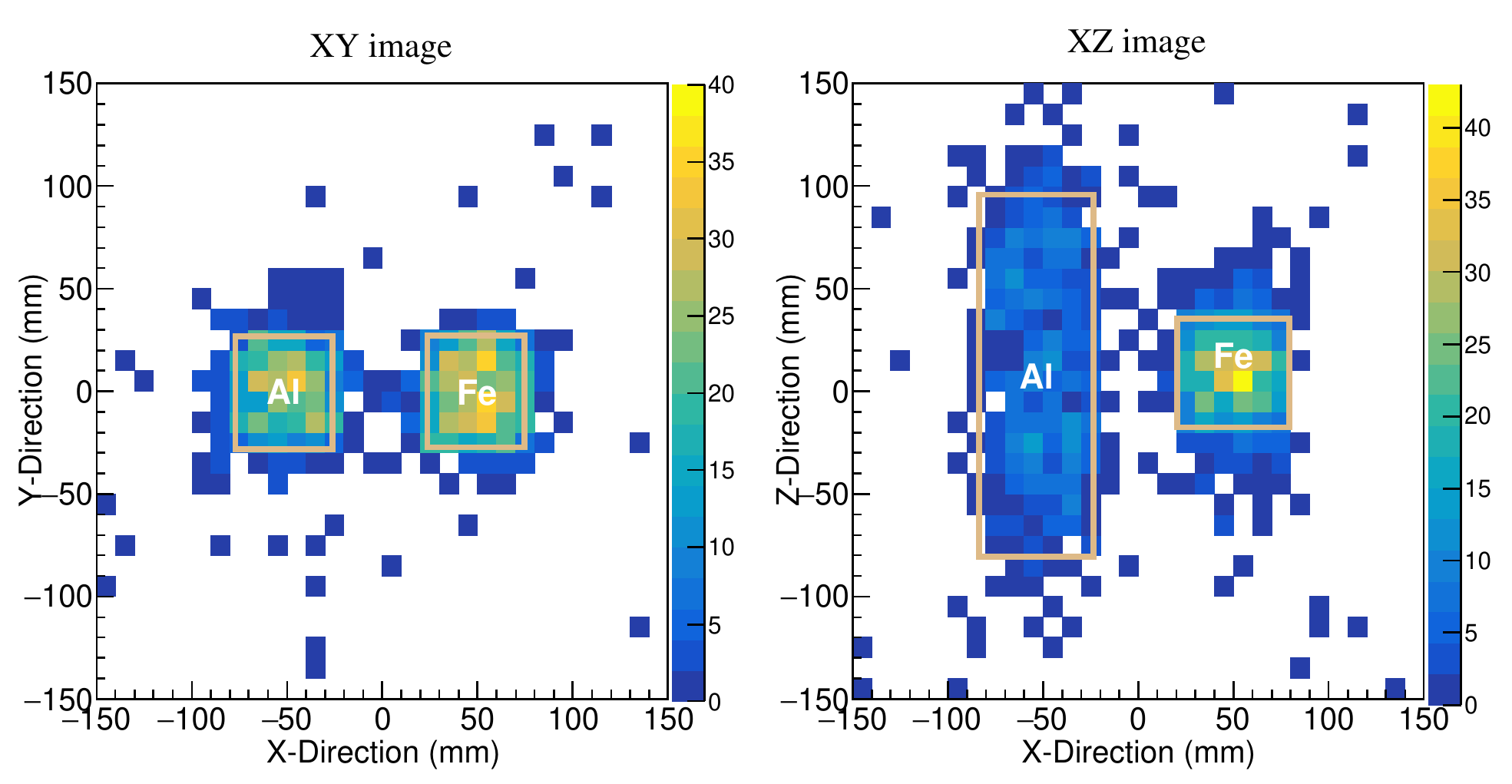}
	\caption{\label{dimen} Images obtained in both XY and XZ direction for two different materials obtained for 200~$\mu$m detector resolution. The boundaries are marked by solid lines.}
\end{figure}
In case of two-dimensional imaging of the ROI, the content in the perpendicular direction to the image is lost. Therefore, objects of dissimilar materials with same cross sectional area and different depth can be falsely identified as alike objects. In a test case, a block of aluminium~(5~cm$~\times$~5~cm~$\times$~18~cm) and a block of iron~(5~cm~$\times$~5~cm~$\times$~5~cm) have been placed. The XY images are analogous as shown in Figure~\ref{dimen}, whereas the XZ image reveals the depth of the aluminium block. This implies the importance of considering XZ / YZ images too when the test objects do not have same dimension along the Z-direction. However, for the sake of simplicity, the present work considers all the test objects to be of same dimension in all directions which has been already mentioned. 

The assumptions used in the proposed PRM are the following: i) The muon exposure should be same for the training and test images, ii) the longitudinal dimension of the training and test objects should be same, otherwise scaled to the size and iii) the tracking detectors should have identical spatial resolution.

A cluster finding method, namely DBSCAN has also been used for material discrimination in parallel. It identifies a test data as an eligible cluster if a specified number of data points, is found within a given radius around the test point. To compare it with the PRM, Similar values of clustering parameters have been chosen. Total scattering points inside the filter matrix of a given sample has been considered as the minimum data point, size of the matrix has been considered as the radius for DBSCAN.

   \begin{figure}[htbp]
	\centering 
	\includegraphics[trim = 0 80 0 100, clip, angle = 0,width=.45\textwidth]{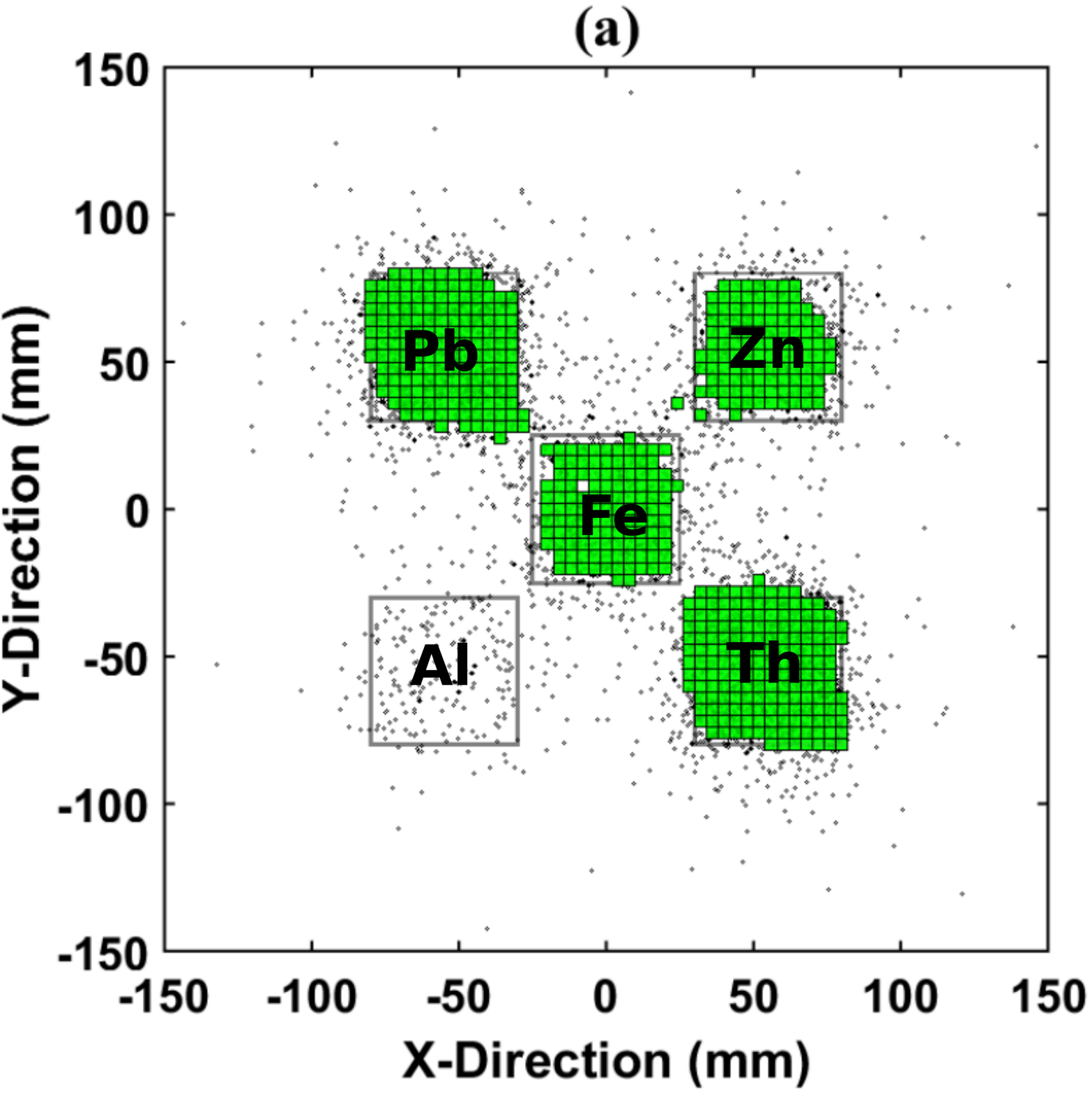}
	\includegraphics[trim = 0 80 0 100, clip, angle = 0,width=.45\textwidth]{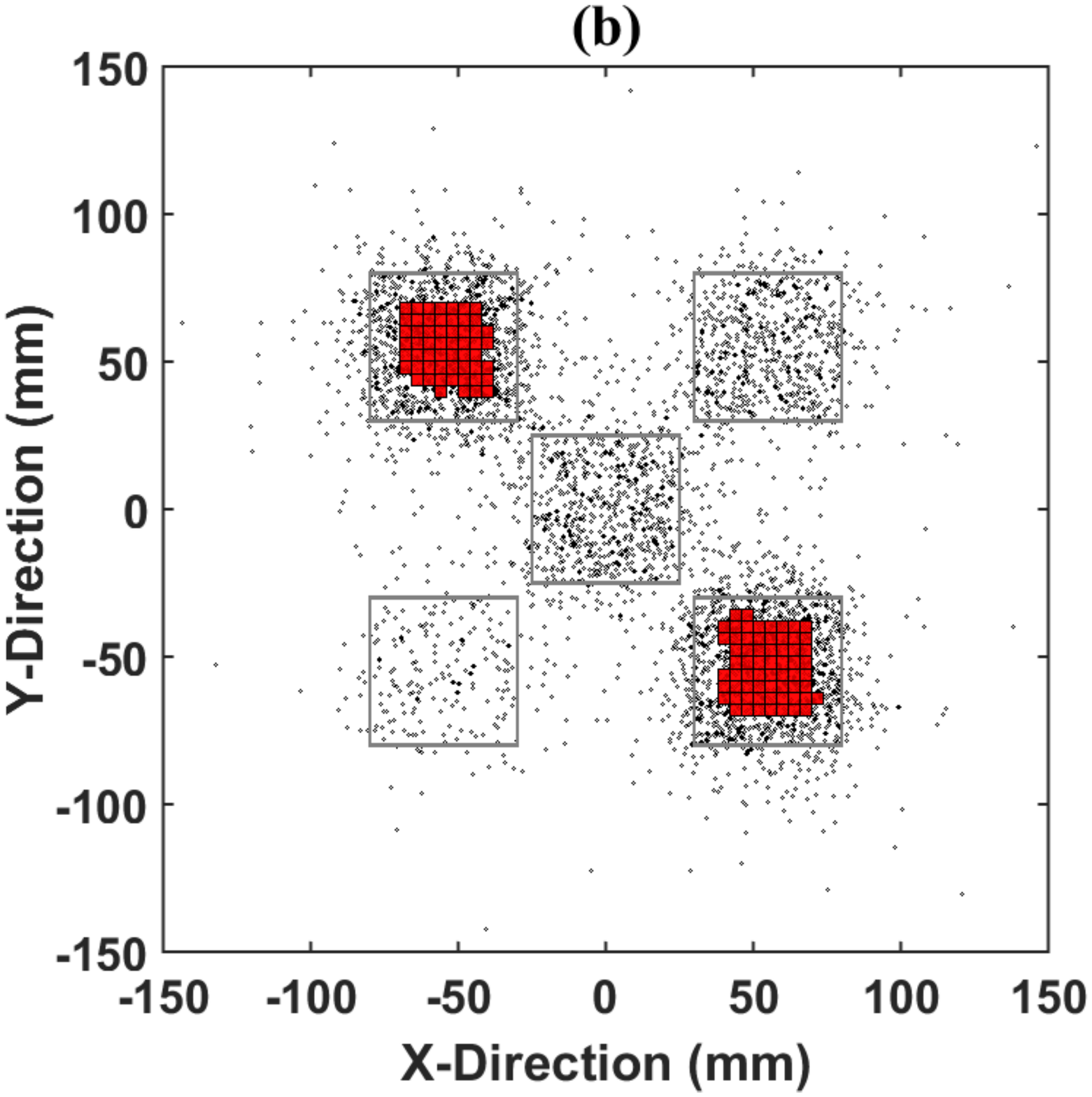}
	\includegraphics[trim = 0 100 0 80, clip, angle = 0,width=.45\textwidth]{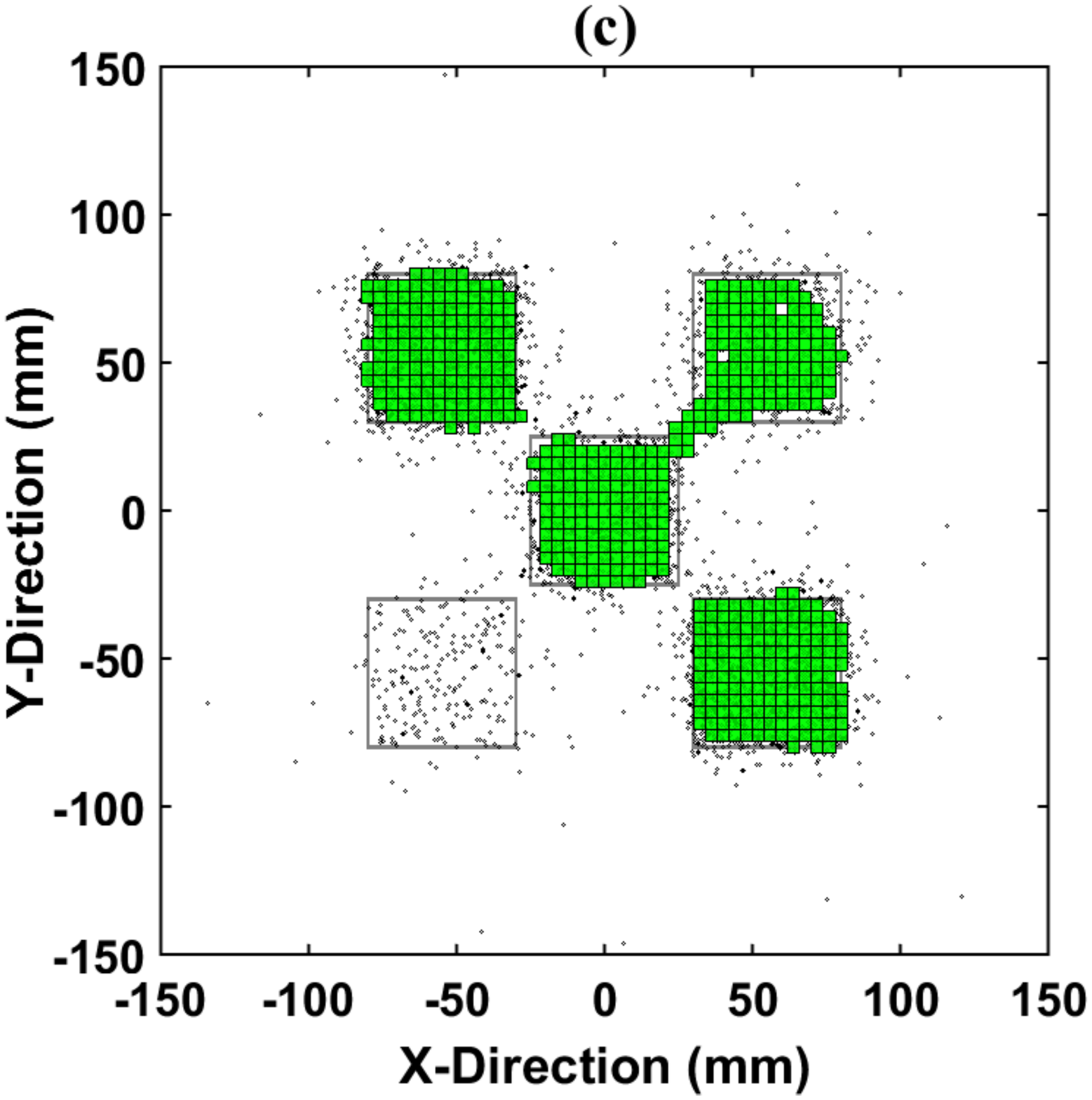}
	\includegraphics[trim = 0 100 0 80, clip, angle = 0,width=.45\textwidth]{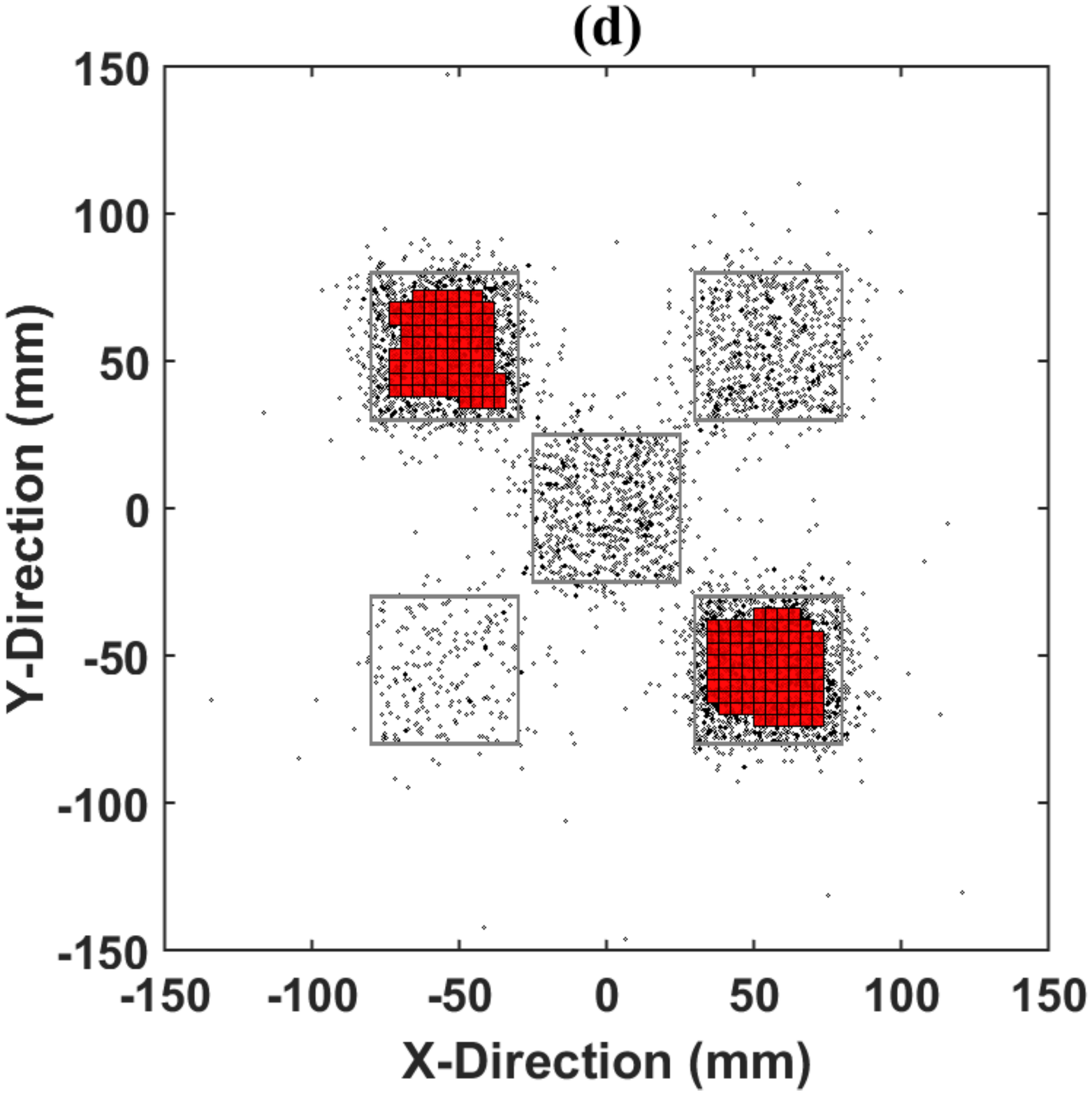}
	
	\caption{\label{prvd_fe} Pattern recognition with (a) iron filter and (b) lead filter with detector resolution of 500 $\mu$m and the same in (c) and (d) with detector resolution of 200 $\mu$m.} 
\end{figure}
\begin{figure}[htbp]
	\centering 
	
	\includegraphics[trim = 0 80 0 100, clip, angle = 0,width=.45\textwidth]{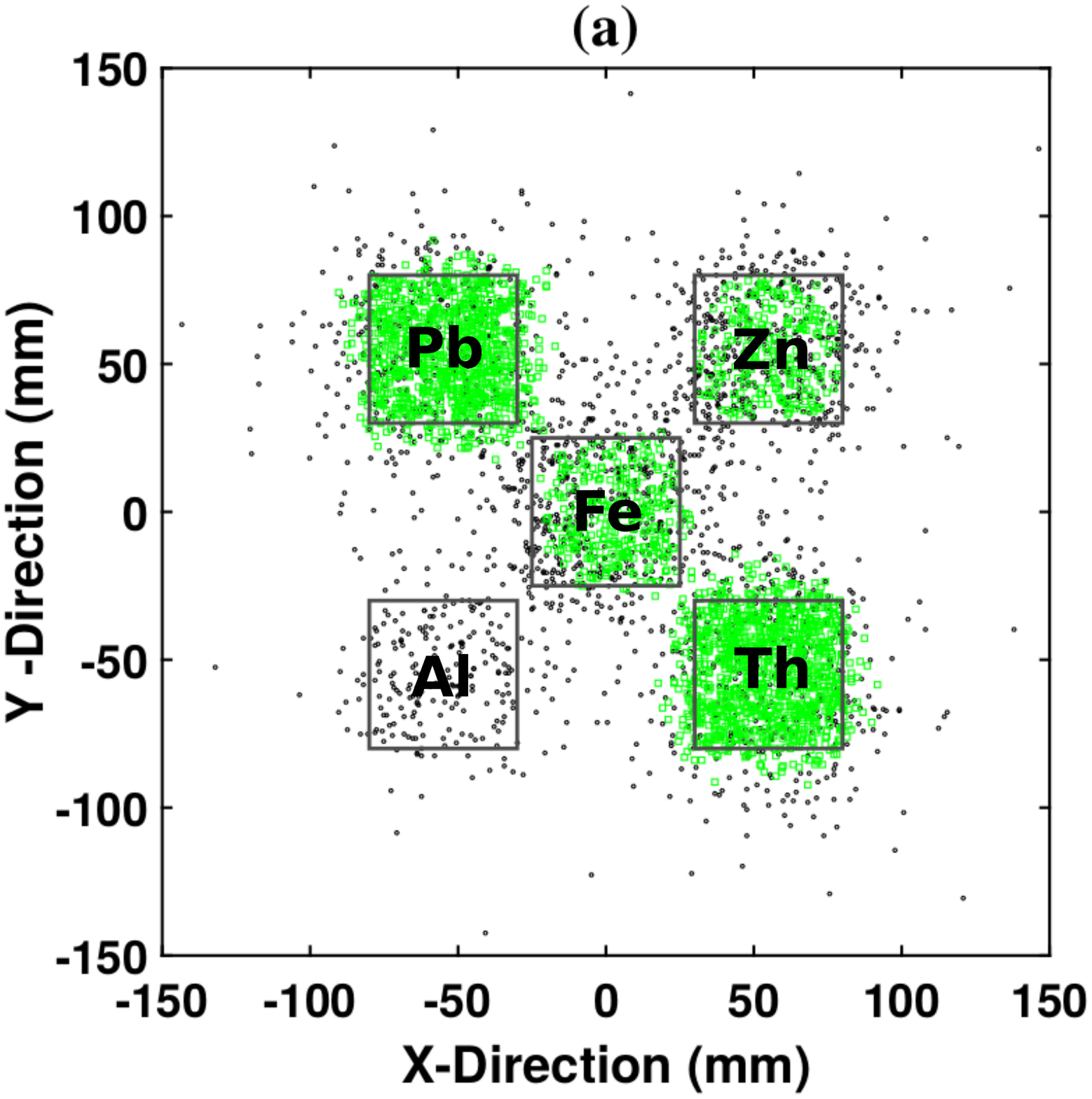}
    \includegraphics[trim = 0 80 0 100, clip, angle = 0,width=.45\textwidth]{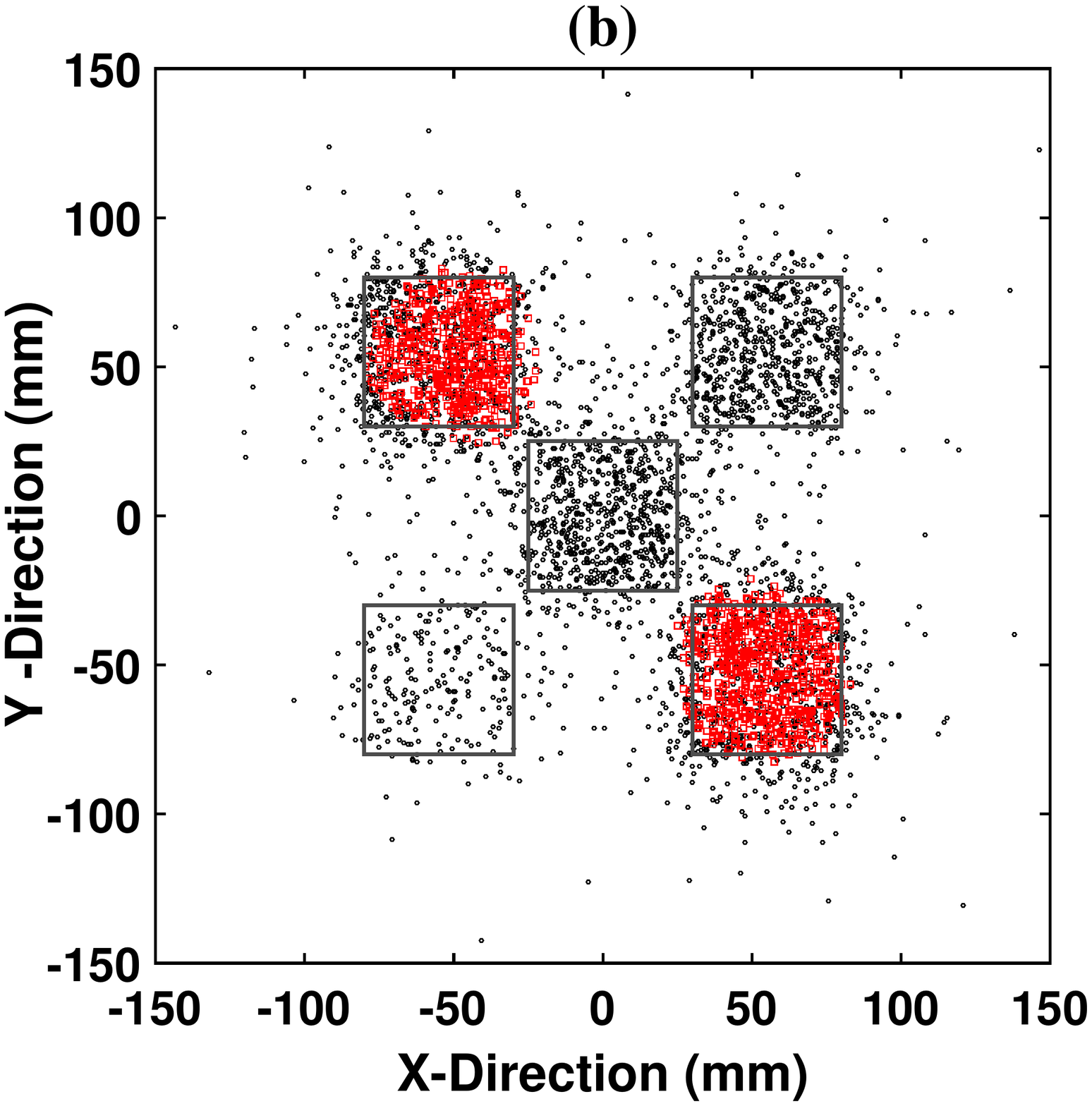}
    \includegraphics[trim = 0 80 0 100, clip, angle = 0,width=.45\textwidth]{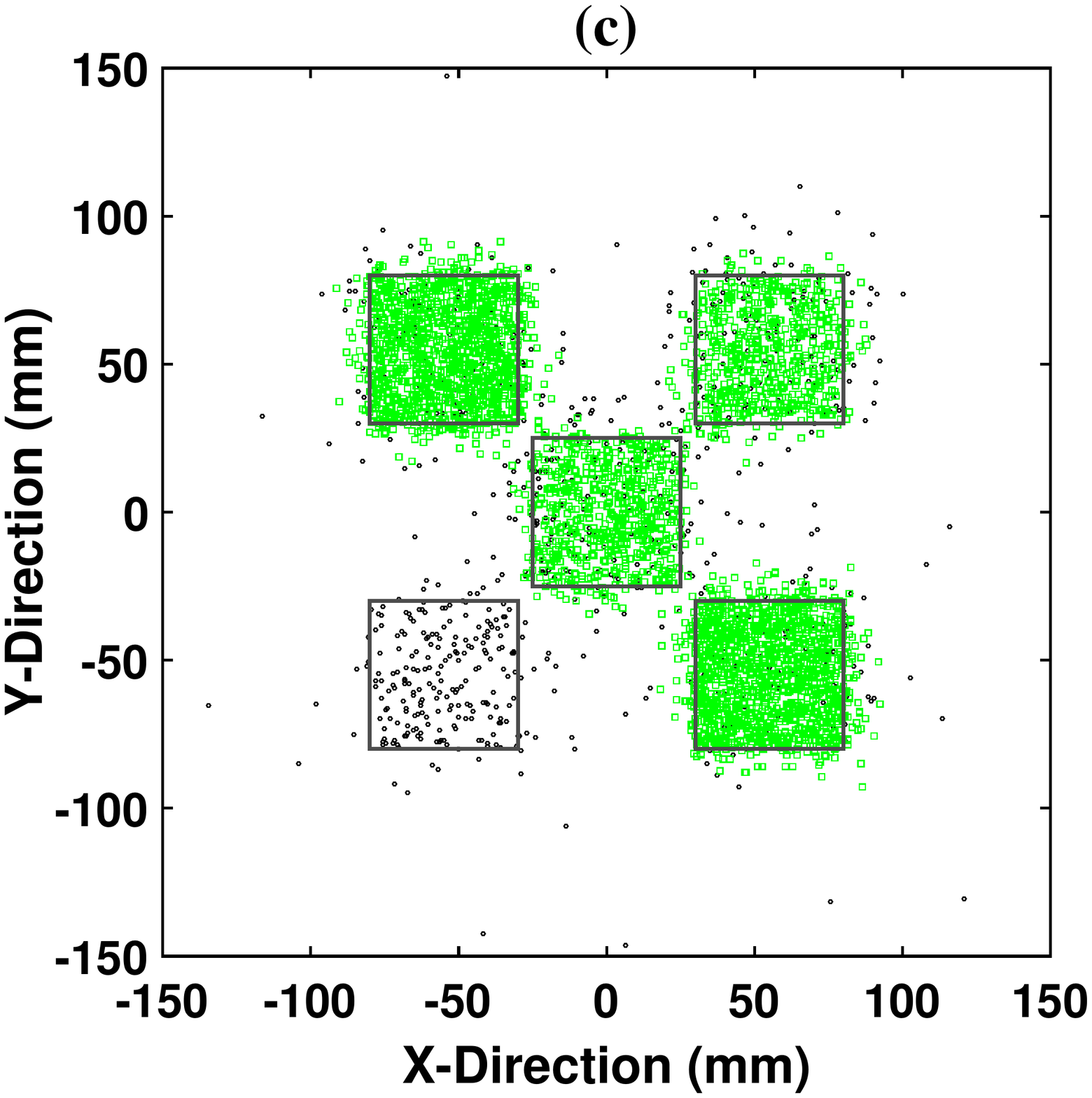}
    \includegraphics[trim = 0 80 0 100, clip, angle = 0,width=.45\textwidth]{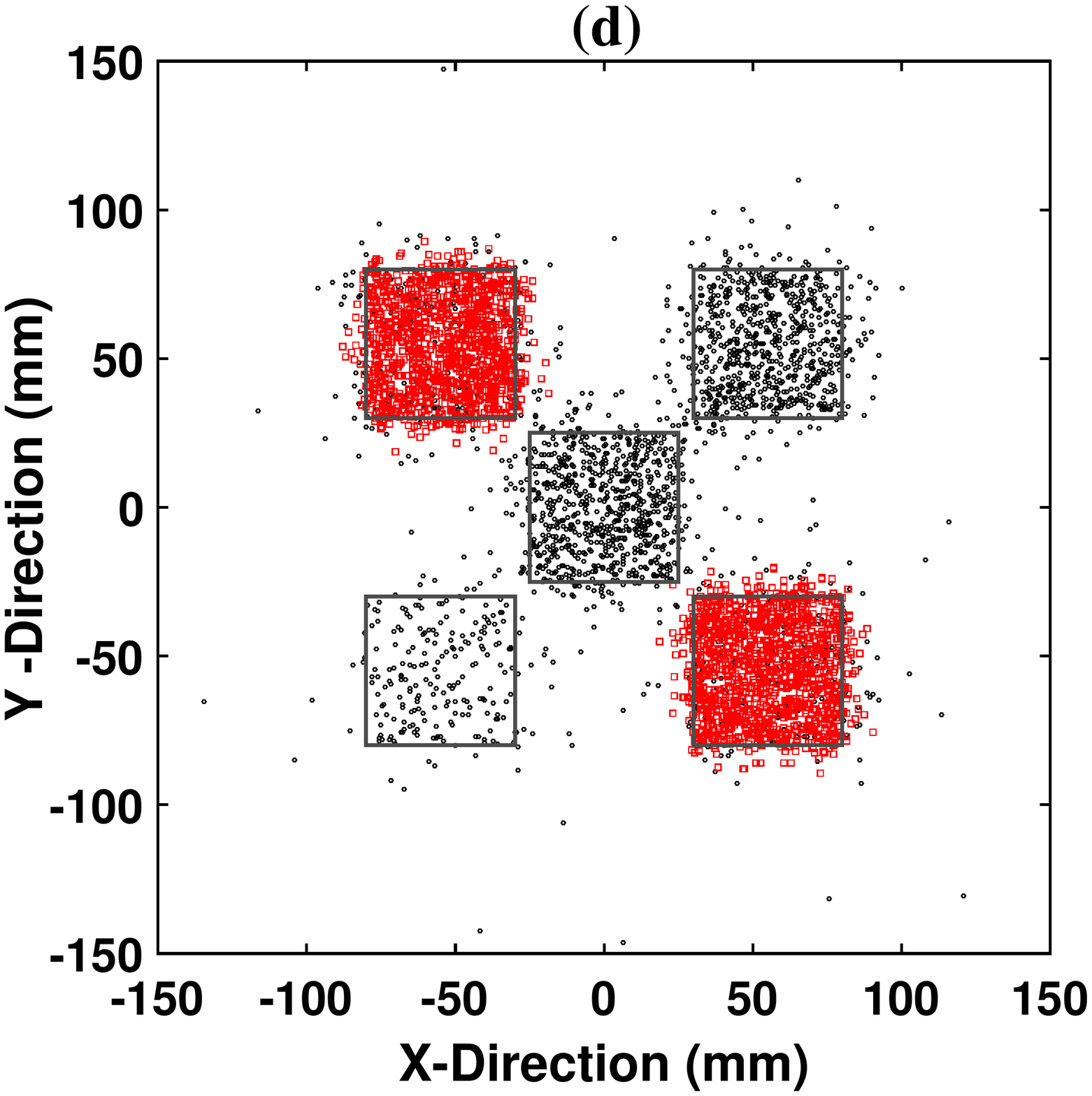}
	
	\caption{\label{DBSCAN_fe} DBSCAN with (a) iron filter and (b) lead filter with detector resolution of 500 $\mu$m and the same in (c) and (d) with detector resolution of 200 $\mu$m.}
\end{figure}
\subsection{Material Identification}
\label{sec:matdis}
The image of a test case with five objects same as that described in section~\ref{sec:ImPer} has been processed with iron and lead filters. The results obtained considering two detector resolutions of 500 and 200 $\mu$m have been shown in figure~\ref{prvd_fe}. The plots for the iron filter and detector resolution 500 and 200 $\mu$m have been displayed in (a) and (c). It can be observed that the iron filter has identified all the materials (iron, zinc, lead, thorium) with $\rho_c$ comparable or larger than that of the iron (green pixels) and rejected the aluminium. The plots produced with the lead filter have been shown in (b) and (d) for 500 and 200 $\mu$m resolution respectively. The figures show that only lead and thorium have been passed by the filter (red pixels) while the rest (aluminium, iron, zinc) have been rejected by the filter. Results for the same case analysed using DBSCAN have been depicted in figure~\ref{DBSCAN_fe}. It can be observed that, the recognition of high-Z material is much better with better spatial resolution of 200 $\mu$m of the detectors.

\begin{figure}[htbp]
	\centering
	\includegraphics[trim = 0 80 0 100, clip, angle = 0,width=.45\textwidth]{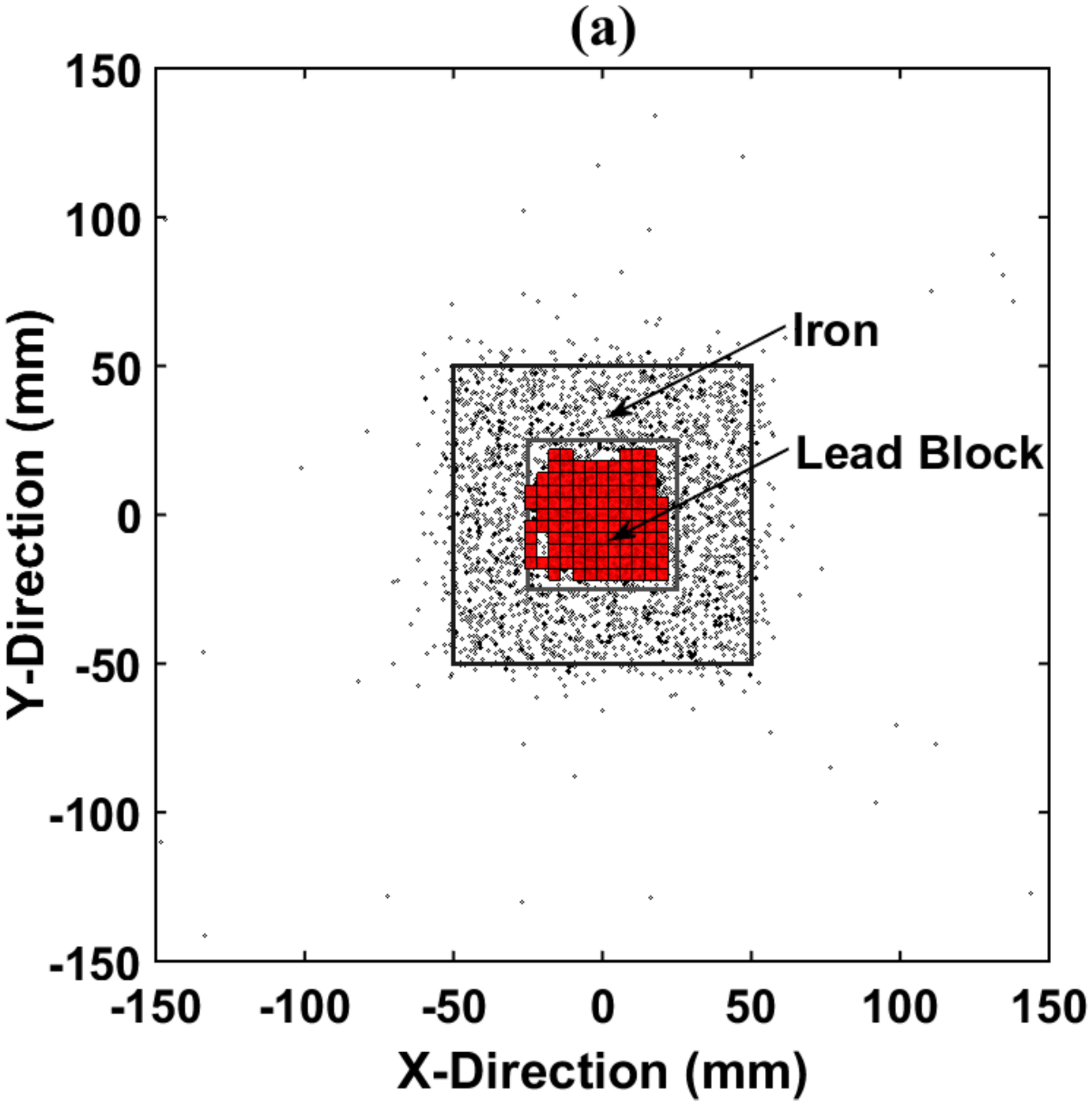}
	\includegraphics[trim = 0 80 0 100, clip, angle = 0,width=.45\textwidth]{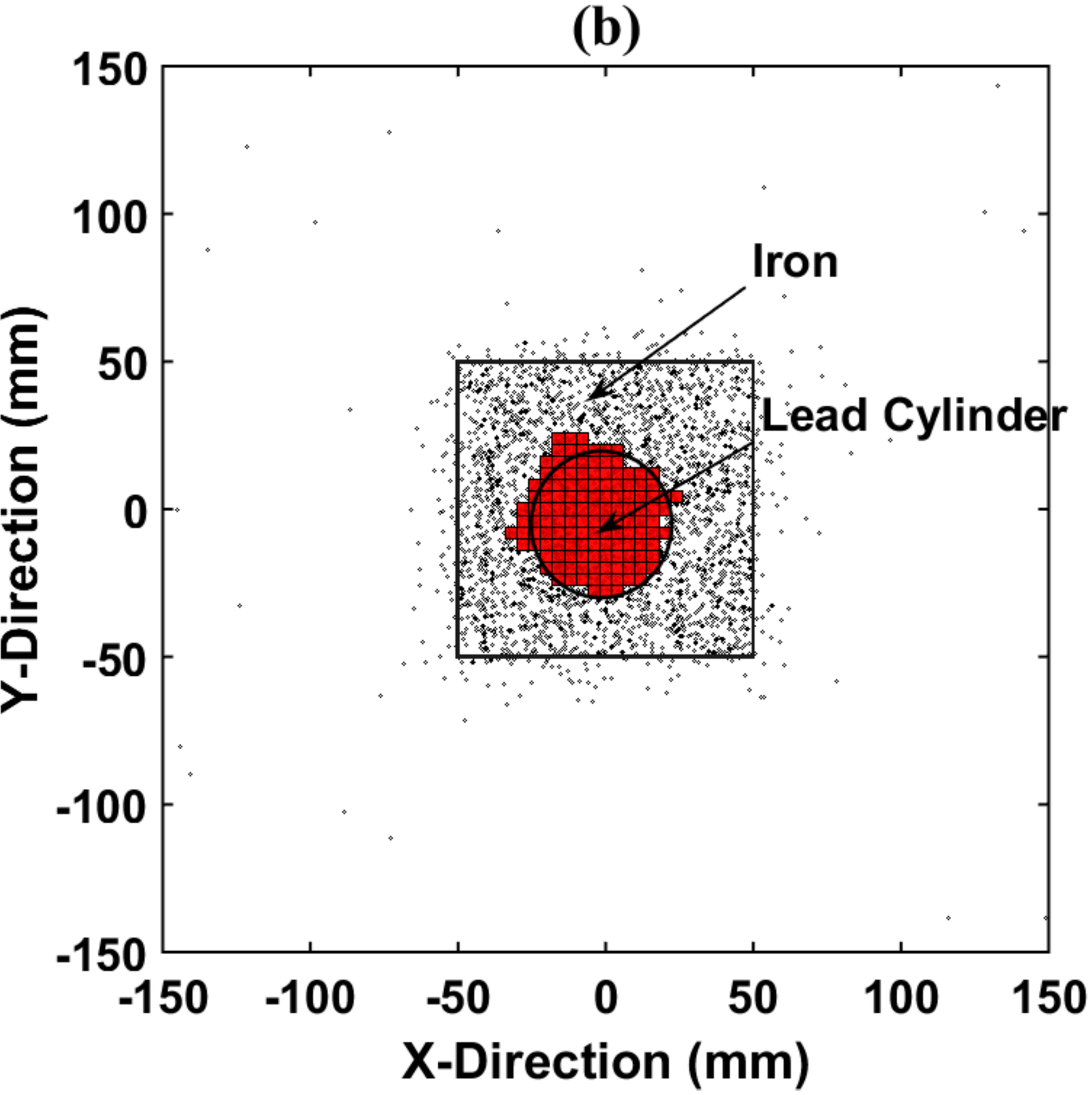}
	\caption{\label{rectrodPR} Results of the PRM with lead filter for the embedded (a) lead cube and (b) lead cylinder for detector resolution of 200 $\mu$m.}
\end{figure}

\begin{figure}[htbp]
	\centering
	\includegraphics[trim = 0 100 0 100, clip, angle = 0,width=.45\textwidth]{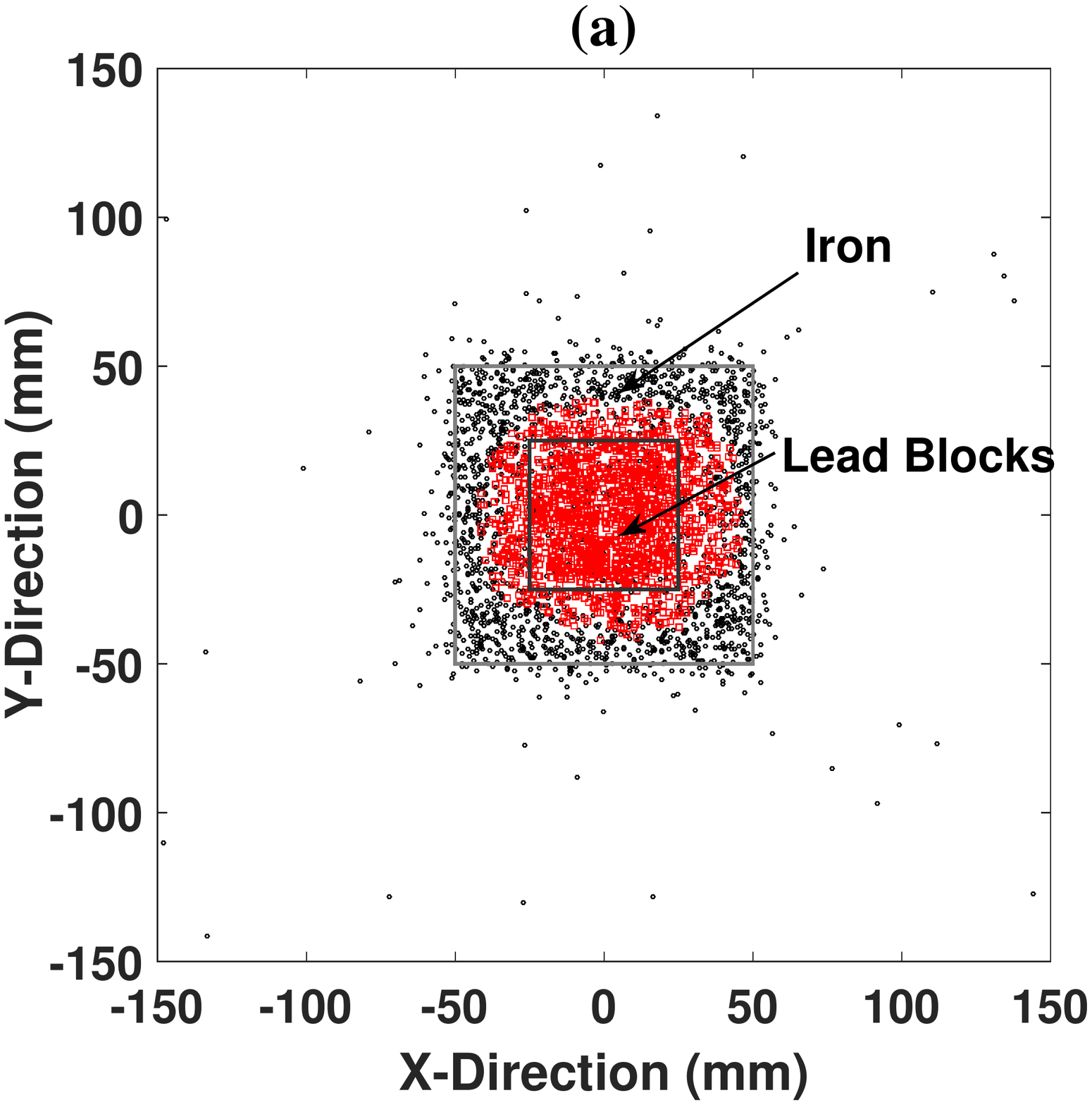}
	\quad
	\includegraphics[trim = 0 100 0 100, clip, angle = 0,width=.45\textwidth]{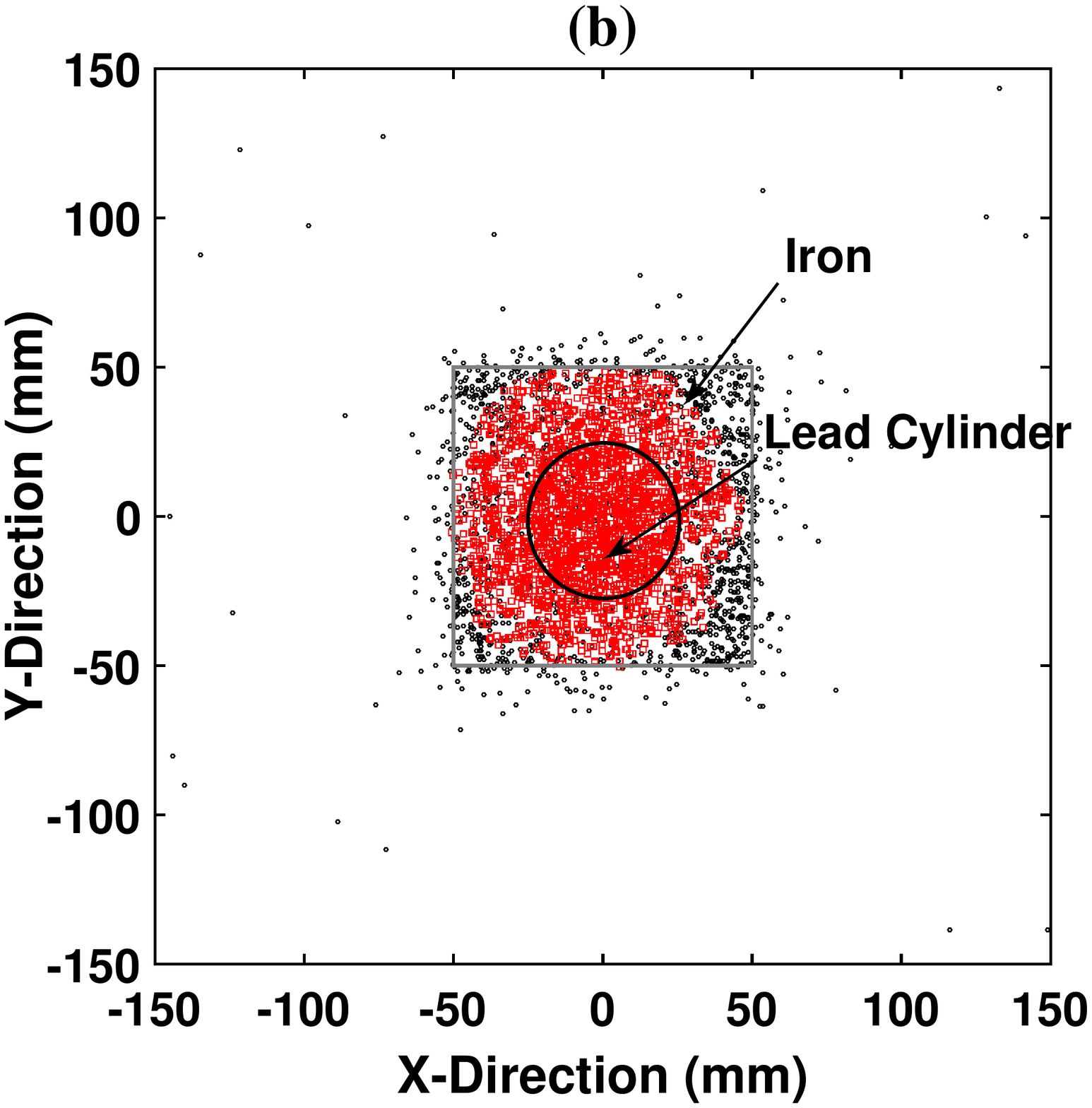}
	\caption{\label{rectrodDB} Results of the DBSCAN with lead filter for the embedded (a) lead cube and (b) lead cylinder for detector resolution of 200 $\mu$m.}
\end{figure}

\subsection{Shape Recognition in Complex Scenarios}
\label{sec:shreco}
To study the performance of the PRM in shape recognition, two test models have been considered. In both the models, an iron block of dimension 10~$\times$~10~$\times$~5~cm$^3$ has contained either of the following two objects embedded at its centre:~(a) a lead cube of side 5 cm, and~(b) a lead cylinder with diameter 5~cm. An improved detector resolution of 200~$\mu$m has been considered here for achieving better definition of the edges of the inner lead cube and the cylinder. For the PRM, the results obtained with the lead filter applied in both the cases have been shown in figure~\ref{rectrodPR} where the actual objects have been marked with solid lines. The images of the same objects when processed with DBSCAN have been displayed in figure~\ref{rectrodDB}. It can be seen from the comparison of the figures that the PRM has been able to distinguish the lead cube and cylinder at the center better than that done by the DBSCAN.

\section{Conclusion}
The performance of a prototype setup has been studied with a simulation framework to accomplish the image formation of a few test objects utilizing the Coulomb scattering of the cosmic muons. In order to reduce computation image processing has been done on two-dimensional images obtained from the projection of reconstructed scattering vertices. The role of the spatial resolution of the detectors in the entire process has been followed in order to optimize the design parameters of the tracking detectors to be used for building the imaging setup. A resolution of the order of 200~$\mu$m has been found to be a good and feasible choice for this purpose. The capability of two-dimensional image processing for discriminating low-Z materials from high-Z materials has qualified the statistical tests with outstanding significance (about~5~$\sigma$). The newly introduced PRM is devised such that, it learns the signature parameters from known samples and identifies materials with similar pattern in test cases and rejects the dissimilar materials. It has been trained with clustering densities of iron and lead and tested in several scenarios. The method has been found to perform satisfactorily for identifying the test objects as low-Z or high-Z materials for reasonable cosmic muon exposure. 
In complex scenarios, where a lead cube~/~cylinder is placed inside a larger iron block the DBSCAN algorithm couldn't identify the transition boundary successfully.
The PRM has been found to be reliable in these scenarios to identify the boundaries. The authors plan to improve the boundary detection capability of PRM in complex scenarios, and extend this work for imaging civil structures and searching for defects with reasonable accuracy. 

\acknowledgments

The author, Sridhar Tripathy, acknowledges the support and cooperation extended by INSPIRE Division, Department of Science and Technology, Govt. of India.

\end{document}